# Optimization of Weighted Individual Energy Efficiencies in Interference Networks

Bho Matthiesen*, Yang Yang† and Eduard A. Jorswieck*
*Institute of Communication Technology, Technische Universität Dresden, Germany
†Interdisciplinary Centre for Security, Reliability and Trust, University of Luxembourg, L-1855 Luxembourg
Email: {bho.matthiesen, eduard.jorswieck}@tu-dresden.de, yang.yang@uni.lu

*Abstract*—This paper studies the maximization of the weighted sum energy efficiency (WSEE). We derive a first-order optimal algorithm applicable to a wide class of communication scenarios exhibiting very fast convergence. We also discuss how to leverage monotonic optimization and fractional programming to obtain a global optimal solution at the cost of higher computational complexity. The WSEE of interference networks is studied in detail with an application to relay-assisted multi-cell communication. This scenario is modeled as a non-regenerative multi-way relay channel and the achievable rate region is derived. We apply the proposed algorithm to this scenario and compare its performance to the global optimal algorithm. The results indicate that the proposed algorithm often achieves the global optimal solution and is close to it otherwise. Convergence is achieved within 10 iterations, while the global optimal solution may require more than $10^6$ iterations.

*Index Terms*—Energy efficiency, resource allocation, interference networks, green communications, 5G networks, fractional programming, monotonic optimization, power control, global optimization, multi-way relay channel, amplify-and-forward

## I. INTRODUCTION

Energy consumption due to information and communication technology already contributes considerably to greenhouse gas emission and operating costs of network operators. In the era of 5G data traffic will increase significantly within the next few years [1]. To avoid exacerbating global warming and to keep the operational expenditures of operators at a reasonable level, the data traffic increase needs to be achieved without a significant increase in energy consumption. Moreover, mobile devices still have very limited energy resources and battery lifetime is a fundamental concern for users. Energy-efficient communication schemes are therefore key technologies in enabling 5G [2], [3].

An important metric often considered is the network-centric global energy efficiency (GEE) defined as the ratio of the total achievable throughput to total consumed energy. In many communication scenarios, e.g. cellular networks with battery powered mobiles and grid powered base stations, a user-centric formulation of the energy efficiency (EE) provides additional insights and might be more suitable for resource allocation.

The work of B. Matthiesen and E. A. Jorswieck is supported by the German Research Foundation (DFG) in the Collaborative Research Center 912 "Highly Adaptive Energy-Efficient Computing". The work of Y. Yang is supported by the ERC project AGNOSTIC, H2020 project SANSA and FNR project SATSENT and PROSAT.
We thank the Center for Information Services and High Performance Computing (ZIH) at TU Dresden for generous allocations of computer time.

One such metric is the weighted sum energy efficiency (WSEE) defined as the weighted sum of each user's individual EE, i.e.,

$$f(\boldsymbol{p}) = \sum_{k=1}^{K} w_k \frac{r_k(\boldsymbol{p})}{\phi_k p_k + P_{c,k}} \tag{1}$$

where $r_k(\boldsymbol{p})$ is the achievable rate of user $k$, $p_k$ the transmit power of user $k$, $\boldsymbol{p} = (p_1, p_2, \ldots, p_K)$, and $w_k$ a positive weight to prioritize some user's EE over others. The positive constants $\phi_k$ and $P_{c,K}$ model the power amplifiers inefficiency and static circuit power consumption of user $k$, respectively.

### A. Problem statement and contributions

We consider maximization of the WSEE over a feasible set $\mathcal{P}$, i.e.,

$$\max_{\boldsymbol{p} \in \mathcal{P}} \sum_{k=1}^{K} w_k \frac{r_k(\boldsymbol{p})}{\phi_k p_k + P_{c,k}}. \tag{P1}$$

For $K = 1$ this is a fractional programming problem and can be solved using well known techniques like Dinkelbach's algorithm [4, Sec. 3.2.1]. For $K > 1$ the problem is known as the sum-of-ratios problem and considerably harder to solve.

Leveraging a recently published successive pseudoconvex approximation framework [5], we propose an algorithm that finds a stationary point of (P1) for $K > 1$ under the constraints that $\mathcal{P}$ is a closed and bounded convex set and $r_k(\boldsymbol{p})$ is concave in $p_k$, a proper function and continuously differentiable on $\mathcal{P}$. Our algorithm exhibits very fast convergence and achieves solutions close to the global optimum. Further, we also obtain the global optimal solution of (P1) utilizing monotonic optimization and fractional programming theory at the price of high computational complexity and slow convergence. In this case, the only requirements on $r_k(\boldsymbol{p})$ and $\mathcal{P}$ are some monotonicity properties that are frequently satisfied in communication systems.

The signal to interference plus noise ratio (SINR)-based rate function $r_k(\boldsymbol{p})$ in (P1) often takes the following form in interference networks:

$$r_k(\boldsymbol{p}) = \log\left(1 + \frac{\theta_k p_k}{\sigma_k^2 + \sum_{j=1}^{K} \eta_{kj} p_j}\right) \tag{2}$$

where the constants $\theta_k > 0$, $\eta_{kj} > 0$, $\eta_{kk} \geq 0$, and $\sigma_k^2 > 0$ for all $k, j = 1, \ldots, K$. Note that in the interference terms in (2),

$\eta_{kk}p_k$ accounts for the self-interference. This model has shown to be valid in a number of wireless communication systems, e.g. hardware impaired massive MIMO systems [6], digital subscriber line (DSL) systems [7], and spectrum optimization in multicarrier systems [7], [8]. Additional examples of communication scenarios with rate function (2) are reported in [9]. In Section IV, we show that (2) also occurs in relay assisted intra-cell communication and study numerical properties of the proposed algorithms by means of this application in Section V.

Our main contributions are summarized as follows:

1) We derive a general framework to optimize the WSEE for a broad range of rate functions with guaranteed convergence to a stationary point.
2) The specialization of our algorithm to interference networks with rate function (2) is discussed. We show that in this case the inner optimization problem decomposes into single variable problems with closed-form solution.
3) The global optimal solution of (P1) is analyzed using fractional monotonic programming theory and the transformation proposed in [10]. To the best of our knowledge, this is the first work to study the numerical behavior (e.g. performance, complexity, and scalability) for this approach of solving (P1) globally.
4) Numerical results for relay-assisted multi-cell communication are provided showcasing the merits of the proposed algorithm in terms of performance and complexity compared to the global optimal solution.

*B. Related work*

Energy-efficient resource allocation, especially in terms of the GEE, has received considerable attention over the past few years. An extensive review of results is provided in [4]. More recent publications include [9], [10] where the EE of general interference networks is considered, and [5] where the performance of state-of-the-art EE maximization algorithms for massive MIMO systems is improved significantly. Compared to the GEE less attention has been paid to the WSEE because the corresponding fractional programming problem cannot be globally solved via some (generalized) Dinkelbach type algorithm. In [10] the global optimal solution of (P1) is considered but no numerical results are given. The authors of [11] consider the WSEE in the context of device-to-device underlay cellular networks and propose an algorithm that obtains a suboptimal solution. In [12], energy-efficient resource allocation in downlink OFDMA networks with various metrics is discussed. In this special case, the objective is separable in the power variables allowing for an efficient solution of the KKT conditions. In [13], the WSEE for coordinated multicell multiuser precoding is optimized. The authors present an iterative two-layer algorithm whereas the inner layer is a block coordinate descent algorithm, which may slow down the convergence. Convergence of the objective function value is established, while the convergence to a stationary point is still left open. The authors of [14] also consider the WSEE of a multi-cell multiuser system. They propose centralized and decentralized algorithms based on successive convex approximation to solve this problem. However, convergence of the proposed algorithms is not guaranteed and a number of auxiliary variables is required increasing the computational complexity of the proposed algorithm. Different applications of the sum-of-ratios problem in the context of communications are, e.g., considered in [15] where resource allocation in simultaneous wireless information and power transfer systems treated. The sum-of-ratios problem itself is one of the most difficult fractional programs and subject to ongoing research in the global optimization community [16]–[18].

*Notation:* Logarithms are to the base $e$. The $k$th Euclidean unit vector is denoted as $\boldsymbol{e}_k$, $(a_k)_k = (a_1, a_2, \dots)$ is a column vector, and $\boldsymbol{p}_{-k} = (p_1, \dots, p_{k-1}, p_{k+1}, \dots, p_K)$. For two vectors $\boldsymbol{x}, \boldsymbol{y}$, we say that $\boldsymbol{x} \geq \boldsymbol{y}$ if $x_i \geq y_i$, for all $i$, and $\boldsymbol{x} \leq \boldsymbol{y}$ if $x_i \leq y_i$.

## II. FIRST-ORDER OPTIMAL SOLUTION

We derive a novel successive convex approximation algorithm to obtain a first-order optimal solution of (P1) under the assumptions that $\mathcal{P} \subseteq \mathbb{R}_{\geq 0}^K$ is a closed and bounded convex set, and $r_k(\boldsymbol{p}) : \mathbb{R}^K \to \mathbb{R}$, $k = 1, \dots, K$, is a proper and continuously differentiable function on $\mathcal{P}$. We also assume that $r_k(\boldsymbol{p})$ is concave in $p_k$ but not jointly in $\boldsymbol{p}$.

Consider the following approximate function defined at a given point $\boldsymbol{p}^t$

$$\widetilde{f}(\boldsymbol{p}; \boldsymbol{p}^t) = \sum_{k=1}^{K} \widetilde{r}_k(p_k; \boldsymbol{p}^t). \tag{3}$$

where

$$\widetilde{r}_k(p_k; \boldsymbol{p}^t) = w_k \frac{r_k(p_k, \boldsymbol{p}^t_{-k})}{\phi_k p_k^t + P_{c,k}} + (p_k - p_k^t) \\ \cdot \left( w_k \frac{-\phi_k r_k(\boldsymbol{p}^t)}{(\phi_k p_k^t + P_{c,k})^2} + \sum_{i \neq k} w_i \frac{\frac{\partial}{\partial p_k} r_i(\boldsymbol{p}^t)}{\phi_i p_i^t + P_{c,i}} \right). \tag{4}$$

The component functions $\widetilde{r}_k(\boldsymbol{p}; \boldsymbol{p}^t)$ are constructed in such a way that they preserve the concavity of $r_k(\boldsymbol{p})$ in $p_k$ while linearizing the other terms w.r.t. $p_k$ at $\boldsymbol{p} = \boldsymbol{p}^t$. We also linearize the denominator of the $k$th term to obtain concave approximate functions. Note that $\widetilde{f}(\boldsymbol{p}; \boldsymbol{p}^t)$ is concave in $\boldsymbol{p}$ because each $\widetilde{r}_k(p_k; \boldsymbol{p}^t_{-k})$ is concave in $p_k$ and constant in all other variables. Thus the approximate problem

$$\max_{\boldsymbol{p} \in \mathcal{P}} \widetilde{f}(\boldsymbol{p}; \boldsymbol{p}^t) \tag{P2}$$

is a convex programming problem and can be solved using standard convex optimization tools.

Let $\mathbb{B}\boldsymbol{p}^t$ be an optimal solution of (P2). Then, $\mathbb{B}\boldsymbol{p}^t - \boldsymbol{p}^t$ is an ascent direction of (P1) [5, Prop. 3] and we define the update $\boldsymbol{p}^{t+1} = \boldsymbol{p}^t + \gamma^t(\mathbb{B}\boldsymbol{p}^t - \boldsymbol{p}^t)$. The step size $\gamma^t = \beta^{m_t}$ is determined by the Armijo rule where $m^t$ is the smallest nonnegative integer such that

$$f(\boldsymbol{p}^t + \beta^{m_t}(\mathbb{B}\boldsymbol{p}^t - \boldsymbol{p}^t)) \geq f(\boldsymbol{p}^t) + \alpha\beta^{m_t}\nabla f(\boldsymbol{x}^t)^T(\mathbb{B}\boldsymbol{p}^t - \boldsymbol{p}^t)$$

with $0 < \alpha < 1$ and $0 < \beta < 1$ being scalar constants.[1] The

---

[1]Please refer to [19, Sec. 1.2.1] for more details on the Armijo rule and how to choose $\alpha$ and $\beta$.

gradient of $f(\boldsymbol{p})$ is computed as

$$\nabla f(\boldsymbol{p}) = \left(\sum_{k=1}^{K} \frac{w_k \nabla r_k(\boldsymbol{p})}{\phi_k p_k + P_{c,k}}\right) - \left(\frac{w_k \phi_k r_k(\boldsymbol{p})}{(\phi_k p_k + P_{c,k})^2}\right)_{k=1,\ldots,K}.$$

This procedure finds a stationary point of (P1) and is summarized in Algorithm 1. The convergence is formally stated in Theorem 1 after first discussing an important special case.

---
**Algorithm 1** Successive convex approximation algorithm
---
1: Initialize $t = 0$, $\boldsymbol{p}^0 \in \mathcal{P}$, $\alpha, \beta \in (0,1)$.
2: **repeat**
3: $\quad \mathbb{B}\boldsymbol{p}^t \leftarrow \arg\max_{\boldsymbol{p} \in \mathcal{P}} \sum_{k=1}^K \widetilde{r}_k(p_k; \boldsymbol{p}^t)$
4: $\quad \gamma^t \leftarrow 1$
5: $\quad$ **while** $f(\boldsymbol{p}^t + \gamma^t(\mathbb{B}\boldsymbol{p}^t - \boldsymbol{p}^t)) < f(\boldsymbol{p}^t) +$
$\qquad\qquad\qquad\qquad \alpha\gamma^t \nabla f(\boldsymbol{p}^t)^T(\mathbb{B}\boldsymbol{p}^t - \boldsymbol{p}^t)$ **do**
6: $\qquad \gamma^t \leftarrow \beta\gamma^t$
7: $\quad$ **end while**
8: $\quad \boldsymbol{p}^{t+1} \leftarrow \boldsymbol{p}^t + \gamma^t(\mathbb{B}\boldsymbol{p}^t - \boldsymbol{p}^t)$
9: $\quad t \leftarrow t + 1$
10: **until** convergence.
---

*Remark 1:* If $\mathcal{P}$ is a Cartesian product, i.e. $\mathcal{P} = \mathcal{P}_1 \times \mathcal{P}_2 \times \cdots \times \mathcal{P}_K$, problem (P2) is separable and can be decomposed into scalar subproblems that can be solved in parallel:

$$\max_{p_k \in \mathcal{P}_k} \widetilde{r}_k(p_k; \boldsymbol{p}^t), \quad k = 1, 2, \ldots, K. \tag{P3}$$

Then, line 3 of Algorithm 1 changes to

$$\mathbb{B}_k\boldsymbol{p}^t \leftarrow \arg\max_{p_k \in \mathcal{P}_k} \widetilde{r}_k(p_k; \boldsymbol{p}^t), \quad k = 1, 2, \ldots, K. \tag{5}$$

In that case the best-response nature of the proposed algorithm becomes apparent. Thus, we can expect fast convergence behavior. Moreover, (P3) can often be solved in closed-form.

We now state the convergence result.

*Theorem 1:* Any limit point of $\{\boldsymbol{p}^t\}$ obtained by Algorithm 1 is a stationary point of (P1).

*Proof:* Problem (P1) together with the assumptions made in this section falls within the class of optimization problems considered in [5] and we can adapt [5, Alg. 1] to solve it.

If the approximate function $\widetilde{f}(\boldsymbol{p}; \boldsymbol{p}^t)$ fulfills the following technical conditions, any limit point of $\{\boldsymbol{p}^t\}$ is a stationary point of (P1) [5, Theorem 1]:

1) $\widetilde{f}(\boldsymbol{p}; \boldsymbol{p}^t)$ is pseudoconcave in $\boldsymbol{p}$ for any $\boldsymbol{p}^t \in \mathcal{P}$.
2) $\widetilde{f}(\boldsymbol{p}; \boldsymbol{p}^t)$ is continuously differentiable in $\boldsymbol{p}$ for any $\boldsymbol{y} \in \mathcal{X}$ and continuous in $\boldsymbol{y}$ for any $\boldsymbol{p} \in \mathcal{P}$.
3) $\nabla_{\boldsymbol{p}}\widetilde{f}(\boldsymbol{p}^t; \boldsymbol{p}^t) = \nabla_{\boldsymbol{p}} f(\boldsymbol{p}^t)$.
4) The solution set of (P2) is nonempty in every iteration $t$.
5) Given any convergent subsequence $\{\boldsymbol{p}^t\}_{t \in \mathcal{T}}$ where $\mathcal{T} \subseteq \{1, 2, \ldots\}$, the sequence $\{\arg\max_{\boldsymbol{p} \in \mathcal{P}} \widetilde{f}(\boldsymbol{p}; \boldsymbol{p}^t)\}_{t \in \mathcal{T}}$ is bounded.

Condition 1) is satisfied because every concave function is also pseudoconcave and $\widetilde{f}(\boldsymbol{p}; \boldsymbol{p}^t)$ is concave in $\boldsymbol{p}$. Condition 2) holds because $r_k(\boldsymbol{p})$ is continuously differentiable by assumption. A sufficient condition for 4) and 5) to hold is the boundedness of $\mathcal{P}$ [5].

Condition 3) is equivalent to

$$\frac{\partial}{\partial p_k}\widetilde{f}(\boldsymbol{p}; \boldsymbol{p}^t)\bigg|_{\boldsymbol{p}=\boldsymbol{p}^t} = \frac{\partial}{\partial p_k}f(\boldsymbol{p})\bigg|_{\boldsymbol{p}=\boldsymbol{p}^t}$$

for all $k = 1, \ldots, K$. First, consider

$$\begin{aligned}\frac{\partial}{\partial p_k}\widetilde{f}(\boldsymbol{p}; \boldsymbol{p}^t)\bigg|_{\boldsymbol{p}=\boldsymbol{p}^t} &= \frac{\partial}{\partial p_k}\widetilde{r}_k(p_k; \boldsymbol{p}^t)\bigg|_{\boldsymbol{p}=\boldsymbol{p}^t} \\ &= \left[w_k \frac{\partial}{\partial p_k}\frac{r_k(p_k, \boldsymbol{p}_{-k}^t)}{\phi_k q_k + P_{c,k}} - w_k\frac{\phi_k r_k(\boldsymbol{p}^t)}{(\phi_k q_k + P_{c,k})^2} \right. \\ &\qquad \left. + \sum_{i \neq k} w_i \frac{\frac{\partial}{\partial p_k} r_i(\boldsymbol{p}^t)}{\phi_i q_i + P_{c,i}}\right]_{\boldsymbol{p}=\boldsymbol{p}^t} \\ &= w_k\left(\frac{\frac{\partial}{\partial p_k}r_k(\boldsymbol{p}^t)}{\phi_k q_k + P_{c,k}} - \frac{\phi_k r_k(\boldsymbol{p}^t)}{(\phi_k q_k + P_{c,k})^2}\right) \\ &\quad + \sum_{i \neq k} w_i \frac{\frac{\partial}{\partial p_k} r_i(\boldsymbol{p}^t)}{\phi_i q_i + P_{c,i}}.\end{aligned} \tag{6}$$

And the right-hand side

$$\begin{aligned}\frac{\partial}{\partial p_k}f(\boldsymbol{p}) &= w_k\frac{\partial}{\partial p_k}\left(\frac{r_k(\boldsymbol{p})}{\phi_k p_k + P_{c,k}}\right) + \sum_{i \neq k} w_i \frac{\frac{\partial}{\partial p_k} r_i(\boldsymbol{p})}{\phi_i p_i + P_{c,i}} \\ &= w_k\left(\frac{\frac{\partial}{\partial p_k}r_k(\boldsymbol{p})}{\phi_k p_k + P_{c,k}} - \frac{\phi_k r_k(\boldsymbol{p})}{(\phi_k p_k + P_{c,k})^2}\right) \\ &\quad + \sum_{i \neq k} w_i \frac{\frac{\partial}{\partial p_k} r_i(\boldsymbol{p})}{\phi_i p_i + P_{c,i}}.\end{aligned} \tag{7}$$

Clearly, (6) = (7)$|_{\boldsymbol{p}=\boldsymbol{p}^t}$ and condition 3) is satisfied. ∎

*Remark 2:* The boundedness of $\mathcal{P}$ is only a sufficient condition for convergence. It can be relaxed as long as conditions 4) and 5) above are still satisfied.

## III. GLOBAL OPTIMAL SOLUTION

Monotonic optimization is a framework to find the globally optimal point of an optimization problem of the following form

$$\max_{\boldsymbol{x} \in \mathcal{G} \cap \mathcal{H}} g(\boldsymbol{x}) \tag{P4}$$

where $g(\boldsymbol{x})$ is an increasing function, $\mathcal{G}$ is a normal set, and $\mathcal{H}$ is a conormal set. A function $f : \mathbb{R}^n \mapsto \mathbb{R}$ is called *increasing* on $\mathbb{R}_{\geq 0}^n$ if $g(\boldsymbol{x}) \leq g(\boldsymbol{x}')$ whenever $0 \leq \boldsymbol{x} \leq \boldsymbol{x}'$. A set $\mathcal{G} \subseteq \mathbb{R}_{\geq 0}^n$ is called *normal (conormal)* if for any two vectors $\boldsymbol{x}, \boldsymbol{y} \in \mathbb{R}_{\geq 0}^n$ such that $\boldsymbol{y} \leq \boldsymbol{x}$ ($\boldsymbol{y} \geq \boldsymbol{x}$), if $\boldsymbol{x} \in \mathcal{G}$, then $\boldsymbol{y} \in \mathcal{G}$. Solving non-convex optimization problems can, in general, require examining every point in the feasible set. Utilizing monotonic optimization theory we are able to solve (P4) in a more efficient way but still with exponential complexity.

The optimization problem

$$\max_{\boldsymbol{x} \in \mathcal{G} \cap \mathcal{H}} g^+(\boldsymbol{x}) - g^-(\boldsymbol{x}), \tag{P5}$$

where $g^+(\boldsymbol{x})$ and $g^-(\boldsymbol{x})$ are increasing functions, is easily transformed into the monotonic optimization problem $\max_{(\boldsymbol{x},t) \in \mathcal{D} \cap \mathcal{E}} g^+(\boldsymbol{x}) + t$, where

$$\mathcal{D} = \{(\boldsymbol{x}, t) \,|\, \boldsymbol{x} \in \mathcal{G}, \, 0 \leq t \leq g^-(\boldsymbol{b}) - \min\{g^-(\boldsymbol{x}), g^-(\boldsymbol{0})\}\}$$
$$\mathcal{E} = \{(\boldsymbol{x}, t) \,|\, \boldsymbol{x} \in \mathcal{H}, \, 0 \leq t \leq g^-(\boldsymbol{b}) - g^-(\boldsymbol{0})\}$$

and $\boldsymbol{b}$ is such that $\mathcal{G} \subset [\mathbf{0}, \boldsymbol{b}]$ [20], [21].

It turns out that problem (P1) is a monotonic optimization problem in disguise. We reformulate it and apply fractional programming theory to reveal its hidden monotonicity. Then, we can solve it with global optimality using monotonic optimization algorithms. In this section, the only requirement on (P1) is that $r_k(\boldsymbol{p})$ can be decomposed into two monotonic functions such that $r_k(\boldsymbol{p}) = r_k^+(\boldsymbol{p}) - r_k^-(\boldsymbol{p})$ and $\mathcal{P} = \mathcal{G} \cap \mathcal{H}$. Then, as proposed in [10],

$$f(\boldsymbol{p}) = \frac{\sum_{k=1}^{K} w_k(r_k^+(\boldsymbol{p}) - r_k^-(\boldsymbol{p})) \prod_{i \neq k}(\phi_k p_k + P_{c,k})}{\prod_{k=1}^{K}(\phi_k p_k + P_{c,k})}. \quad (8)$$

Note that both the numerator and the denominator are differences of increasing functions in $\boldsymbol{p}$ because the product of nonnegative increasing functions is also increasing.

We know from fractional programming theory that the global maximum of (8) can be found with Dinkelbach's algorithm [22] if

$$\max_{\boldsymbol{p} \in \mathcal{P}} F(\boldsymbol{p}; \lambda) \quad (P6)$$

can be solved with global optimality, where

$$F(\boldsymbol{p}; \lambda) = \sum_{k=1}^{K} \left( w_k(r_k^+(\boldsymbol{p}) - r_k^-(\boldsymbol{p})) \prod_{i \neq k}(\phi_k p_k + P_{c,k}) \right)$$
$$- \lambda \left( \prod_{k=1}^{K}(\phi_k p_k + P_{c,k}) \right).$$

This can be achieved using monotonic optimization since (P6) is an instance of (P5). The procedure is stated formally in Algorithm 2. Please refer to [10] for more details on the

---

**Algorithm 2** Fractional monotonic programming algorithm

Initialize $\varepsilon > 0$, $t = 0$, $\boldsymbol{p}^0 \in \mathcal{P}$
**repeat**
$\quad t \leftarrow t + 1$
$\quad \lambda^t \leftarrow f(\boldsymbol{p}^{t-1})$
$\quad \boldsymbol{p}^t \leftarrow \arg\max_{\boldsymbol{p} \in \mathcal{P}} F(\boldsymbol{p}; \lambda^t)$
**until** $F(\boldsymbol{p}^t; \lambda^t) \leq \varepsilon$

---

outlined solution method in this section, to [20] and [21] for more details on monotonic optimization theory and algorithms solving (P4), and to [4] for more details on Dinkelbach's algorithm and fractional programming in general.

## IV. APPLICATION TO INTERFERENCE NETWORKS

Consider maximizing the WSEE in Gaussian interference networks with rate function (2) and average transmit power constraints $\bar{P}_k$ at user $k$. It is easily verified from its second derivatives that (2) is concave in $p_k$ on $\mathbb{R}_{\geq 0}$ and convex in all other variables. Moreover, it is a proper function and continuously differentiable. The feasible set $\mathcal{P} = [0, \bar{P}_1] \times \cdots \times [0, \bar{P}_K]$ is bounded, closed and convex. Hence, all technical properties listed in Section II are satisfied and we can apply Algorithm 1. The subproblems encountered in Algorithm 1 are separable and can be solved in parallel since $\mathcal{P}$ is a Cartesian product (cf. Remark 1). More specifically, the optimization problems to solve in each iteration are

$$\max_{0 \leq p_k \leq \bar{P}_k} \widetilde{r}_k(p_k; \boldsymbol{p}^t),$$

for $k = 1, 2, \ldots, K$. The approximate function $\widetilde{r}_k(p_k; \boldsymbol{p}^t)$ has the structure

$$\widetilde{r}_k(p_k; \boldsymbol{p}^t) = ar_k(p_k; \boldsymbol{p}_{-k}^t) + bp_k + c$$

with $r_k = \log(1 + \frac{\theta_k p_k}{\eta_{kk} p_k + d})$ and $a$, $b$, $c$, and $d$ being constants depending on $\boldsymbol{p}^t$ and $\boldsymbol{p}_{-k}^t$, respectively. A closed-form solution is easily obtained by setting the first derivative of $\widetilde{r}_k(p_k; \boldsymbol{p}^t)$ to zero. Thus, Algorithm 1 has very low complexity. To construct $\widetilde{r}_k(p_k; \boldsymbol{p}^t)$ and $\nabla f(\boldsymbol{p})$ we require the gradient of $r_k(\boldsymbol{p})$ which is easily computed as

$$\nabla r_k(\boldsymbol{p}) = \frac{\theta_k}{\theta_k p_k + \sigma_k^2 + \sum_{j=1}^{K} \eta_{kj} p_j}$$
$$\left( \boldsymbol{e}_k - \left( \frac{\eta_{ki} p_k}{\sigma_k^2 + \sum_{j=1}^{K} \eta_{kj} p_j} \right)_{i=1,\ldots,K} \right).$$

As already indicated in the introduction, this problem can be solved with global optimality applying the results in Section III. To verify that $r_k(\boldsymbol{p})$ is indeed the difference of two increasing functions, note that $r_k(\boldsymbol{p}) = r_k^+(\boldsymbol{p}) - r_k^-(\boldsymbol{p})$ with $r_k^+(\boldsymbol{p}) = \log\left(\sigma_k^2 + \theta_k p_k + \sum_{j=1}^{K} \eta_{kj} p_j\right)$ and $r_k^-(\boldsymbol{p}) = \log\left(\sigma_k^2 + \sum_{j=1}^{K} \eta_{kj} p_j\right)$. Further, $\mathcal{P}$ also satisfies the requirements since $\mathcal{P} = \mathcal{G} \cap \mathcal{H}$ with $\mathcal{G} = \{\boldsymbol{p} : p_k \leq \bar{P}_k\}$ and $\mathcal{H} = \{\boldsymbol{p} : p_k \geq 0\}$ being normal and conormal sets, respectively.

### A. Non-Regenerative Multi-Way Relay Channels

As a practical example, we consider relay-assisted multi-cell communication with multiple unicast transmissions as illustrated in Fig. 1. In this scenario $K$ users near the cell edges want to communicate with each other in multiple unicast transmissions. They communicate over a layer-1 relay located near the cell edges instead of utilizing the connections to their respective base stations which, e.g., might suffer from high attenuation, near-far effects, or high spectrum utilization due to other users.

This is an instance of a multi-way relay channel (MWRC) with amplify-and-forward (AF) relaying. We assume single user receivers at each node, i.e. interference is treated as noise. The multiple unicast transmissions take place in a circular fashion, i.e. user 1 transmits to user 2, user 2 to user 3, ..., and user $K$ to user 1. Each node has a single antenna, an average transmit power constraint, and suffers from additive white Gaussian noise.

We optimize the individual EEs of the users while assuming that the relay transmits at maximum power. In the context of the multi-cell scenario in Fig. 1 this approach is reasonable since the mobile user terminals are battery powered and the relay is most likely connected to the power grid. The weights $w_k$, $k = 1, 2, \ldots, K$, serve to prioritize the EEs of individual users over others, e.g. to deal with low battery levels.

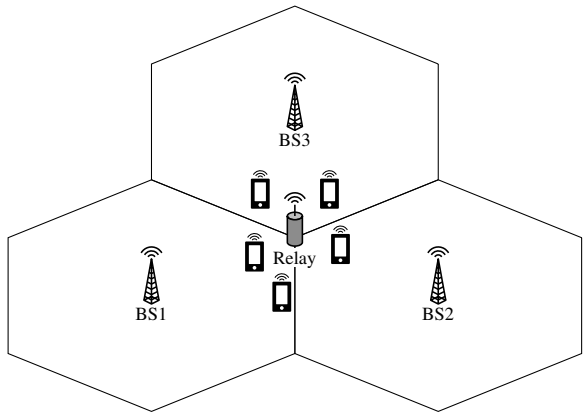

Fig. 1. Relay-assisted multi-cell communication.

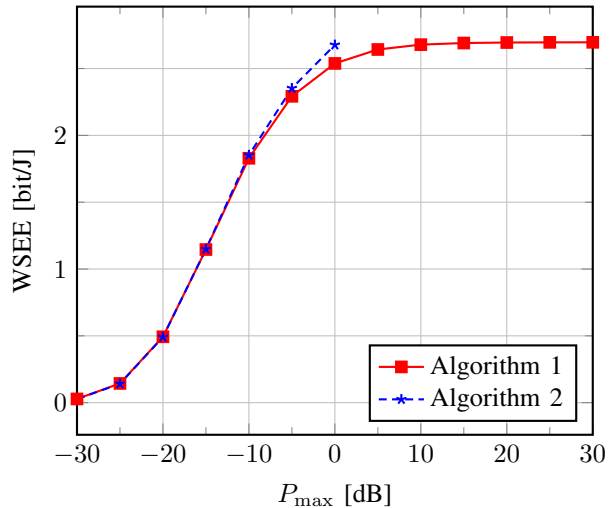

Fig. 2. Weighted sum energy efficiency (WSEE) of relay-assisted multi-cell communication modeled as a non-regenerative multi-way relay channel. Results are averaged over 664 i.i.d. channel realizations. The global optimal solution of Algorithm 2 is compared to the solution of Algorithm 1 satisfying first-order optimality conditions.

The achievable rate region for an average relay power constraint $P_0$ and transmit power $p_k$ at user $k$ is given in the following lemma. The channel from user $k$ to the relay is denoted as $h_k$ and the reverse channel as $g_k$, while user $k$'s and the relay's noise powers are $N_k$ and $N_0$, respectively. For notational convenience, we define $K + 1 \mapsto 1$.

*Lemma 1:* A rate tuple $(R_1, R_2, \ldots, R_K)$ is achievable for the Gaussian MWRC with AF relaying and treating interference as noise at the decoders if

$$r_k(\boldsymbol{p}) < \mathrm{C}\left(\frac{|h_k|^2 p_k}{N_0 + \sum_{\substack{i \neq k \\ i \neq k+1}} |h_i|^2 p_i + \widetilde{g}_{k+1}^{-1}(N_0 + \sum_{i=1}^K |h_i|^2 p_i)}\right),$$

where $\widetilde{g}_k = |g_k|^2 \frac{P_0}{N_k}$ and $\mathrm{C}(x) = \log(1+x)$.

*Proof sketch:* A straightforward extension of [23, Corollary 1] yields the general $K$ user rate region for discrete, memoryless channels. Then adapt this result to Gaussian channels using the standard procedure [24, Chapter 3]. Apply this to the channel defined above with $\mathbb{E}[X_k^2] = p_k$, $k = 0, 1, \ldots, K$. Since the partial derivative of each $R_k$ with respect to $p_0$ is always nonnegative, $R_k$ is an increasing function of $p_0$ and $p_0 = P_0$ is throughput-optimal. ∎

The rate functions in Lemma 1 are instances of (2) and we can apply Algorithms 1 and 2 to optimize the weighted sum of the individual EEs. We present numerical results for this application in the next section.

## V. NUMERICAL EVALUATION

We optimize the WSEE of the multi-cell scenario discussed in Section IV-A in order to compare the performance of Algorithm 1 to the global optimal solution obtained by Algorithm 2 and discuss some general properties of the optimal WSEE. All nodes have the same transmit power constraints, power consumption, and noise variance, i.e., $P_0 = \bar{P}_k = P_{\max}$, $P_{c,k} = 1$, $\phi_k = 2.5$, and $N_0 = N_k = 10^{-2}$, for $k = 1, 2, \ldots, K$. Channels are assumed reciprocal and chosen independent and identically distributed (i.i.d.) with circularly symmetric complex Gaussian distribution, i.e., $h_k \sim \mathcal{CN}(0,1)$ and $g_k = h_k^*$. The maximum transmit power $P_{\max}$ is stepped from -30 dB to 30 dB.

Results for $K = 3$ users are reported in Fig. 2. The WSEE is an increasing function up to around $P_{\max} = 15$ dB where it saturates. From this point onward the optimal transmit power stays constant and increasing $P_{\max}$ does not result in higher throughput. Algorithm 2[2] yields a global optimal solution within a given tolerance at the price of very high computational complexity. Instead, Algorithm 1 obtains a first order optimal solution which is apparently globally optimal for small signal-to-noise ratios (SNRs). For higher SNRs, however, Algorithm 2 yields slightly better results since saturation of the WSEE occurs a bit earlier in Algorithm 1. Due to the high computational complexity of Algorithm 2 and the increasing number of required iterations for convergence at higher $P_{\max}$, Algorithm 2 did not terminate within reasonable time for $P_{\max} > 0$ dB. Moreover, Algorithm 2 converged only for 664 out of 1,000 i.i.d. channel realizations for $P_{\max} \leq 0$ dB. Thus, for both algorithms, we included only results from these 664 channel realizations.

The average number of iterations required for convergence is reported in Table I for both algorithms. Each iteration of Algorithm 2 requires the solution of a monotonic optimization problem. Since this inner problem defines the complexity of Algorithm 2 we also report the required iterations for the solution of one such subproblem. The total number of iterations is listed in the last column. It can be observed that the number of iterations for Algorithm 2 grows strongly with $P_{\max}$ due to the growing feasible set. One important thing to note here is that in the solution of the inner problem, with each iteration the number of candidate points increases. Hence, the required time per iteration increases with the iteration count.

Instead, Algorithm 1 requires very few iterations for conver-gence and the variable update in each iteration has a closed-form expression. We implemented the computation of Fig. 2 such that the optimal solution of the previous $P_{\max}$ step is used as starting point in the next step.[3] This explains the decrease of required iterations in the saturation region. As initial starting point for each channel realization we used maximum transmit power at all nodes.

---

[2]The inner optimization problem in Algorithm 2 is solved with the Polyblock algorithm [20].

TABLE I
AVERAGE NUMBER OF ITERATIONS FOR THE PROPOSED ALGORITHMS.

| $P_{\max}$ [dB] | Algorithm 1 | Algorithm 2 | | |
|---|---|---|---|---|
| | | Outer | Inner | Total |
| -30 | 2.44 | 1.11 | 2,282 | 2,543 |
| -25 | 4.97 | 1.66 | 7,801 | 12,981 |
| -20 | 6.49 | 1.80 | 13,837 | 24,841 |
| -15 | 6.87 | 1.67 | 19,987 | 33,472 |
| -10 | 9.97 | 2.00 | 57,086 | 114,258 |
| -5 | 3.27 | 3.05 | 202,659 | 617,134 |
| 0 | 5.25 | 4.02 | 634,531 | 2,547,681 |
| 5 | 3.94 | – | – | – |
| 10 | 4.51 | – | – | – |
| 15 | 3.84 | – | – | – |
| 20 | 2.83 | – | – | – |
| 25 | 2.49 | – | – | – |
| 30 | 2.36 | – | – | – |

*A. Complexity*

Algorithm 1 scales very well with increasing number of users. While no rigorous complexity analysis is available for the successive pseudoconvex approximation framework [5] yet, experience shows that the number of required iterations does not increase much. The numerical complexity of each iteration increases linearly with the number of users for a separable inner problem and is polynomial otherwise due to the inner problem being a convex optimization problem.

Instead, the complexity of Algorithm 2 grows very fast with the number of users. The inner problem has exponential complexity while the outer algorithm has superlinear convergence. Thus, the overall complexity of Algorithm 2 is exponential in the number of users. From the numerical results above it can be observed that even the 3-user case is barely treatable for some SNR values.

## VI. CONCLUSION

In this paper, we propose an iterative algorithm that yields a first-order optimal solution of the WSEE maximization problem for general interference networks under very modest requirements on the user's rate functions. Numeric evidence shows that the obtained solution is close to the global optimal solution obtained by fractional monotonic programming. The data also clearly shows the very high computational complexity of using fractional monotonic programming to solve the sum-of-ratios problem. This suggest that future research on the WSEE problem should not only consider other suboptimal, low complexity algorithms but also other global optimization methods tailored to the sum-of-ratios problem.

---

[3]The same approach is applicable to Algorithm 2. However, due to its high run time we ran all steps in parallel.